\documentclass{article}

\usepackage{arxiv}

\usepackage[utf8]{inputenc} 
\usepackage[T1]{fontenc}    
\usepackage{hyperref}       
\usepackage{url}            
\usepackage{booktabs}       
\usepackage{amsfonts}       
\usepackage{nicefrac}       
\usepackage{microtype}      
\usepackage{lipsum}		
\usepackage{graphicx}
\usepackage{natbib}
\usepackage{doi}
\usepackage{tikz}

\usetikzlibrary{shapes.geometric, arrows}
\usetikzlibrary{positioning}

\tikzstyle{startstop} = [rectangle, rounded corners, minimum width=3cm, minimum height=1cm,text centered,
text width=3cm,  draw=black, fill=red!30]
\tikzstyle{io} = [trapezium, trapezium left angle=70, trapezium right angle=110, minimum width=1cm,
text width=3cm,  minimum height=1cm, text centered, draw=black, fill=blue!30]
\tikzstyle{process} = [rectangle, minimum width=3cm,
text width=3cm,  minimum height=1cm, text centered, draw=black, fill=orange!30]
\tikzstyle{decision} = [diamond, minimum width=3cm,
text width=3cm,  minimum height=1cm, text centered, draw=black, fill=green!30]
\tikzstyle{arrow} = [thick,->,>=stealth]
\tikzstyle{request} = [rectangle, rounded corners, minimum width=3cm, minimum height=1cm,text centered,
text width=3cm,  draw=black, fill=green!30]

\title{4TCT \\ A 4chan Text Collection Tool}

\author{ \href{https://orcid.org/0009-0000-1581-4021}{\includegraphics[scale=0.06]{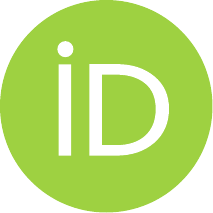}\hspace{1mm}Jack H. Culbert}\thanks{\href{https://www.gesis.org/en/institute/staff/person/John.Culbert}{GESIS Personal Webpage}} \\
	GESIS -- Leibniz Institute for the Social Sciences\\
	Unter Sachsenhausen 6-8 \\
    50667 Köln \\
	\texttt{jack.culbert@gesis.org}
}

\renewcommand{\headeright}{Technical Report}
\renewcommand{\undertitle}{Technical Report}
\renewcommand{\shorttitle}{4TCT}

\hypersetup{
pdftitle={4TCT, A 4chan Text Collection Tool},
pdfsubject={cs.DL},
pdfauthor={Jack H. Culbert,},
pdfkeywords={4chan, Data Collection Tool, GESIS, Sociology, Python, Docker},
}

\begin{document}
\maketitle

\begin{abstract}
    4chan is a popular online imageboard which has been widely studied due to an observed concentration of far-right, antisemitic, racist, misogynistic, and otherwise hateful material being posted to the site, as well as the emergence of political movements and the evolution of memes which are posted there, discussed in Section \ref{sec:sociology}. We have created a tool developed in Python which utilises the 4chan API to collect data from a selection of boards. This paper accompanies the release of the code via the github repository: \href{https://github.com/jhculb/4TCT}{https://github.com/jhculb/4TCT}. We believe this tool will be of use to academics studying 4chan by providing a tool for collection of data from 4chan to sociological researchers, and potentially contributing to GESIS' Digital Behavioural Data project.
\end{abstract}


\section*{Disclaimer}
\label{sec:disclaimer}
Both the author and GESIS are not affiliated, associated, authorized, endorsed by, or in any way officially connected with 4chan, or any of its subsidiaries or its affiliates. The official 4chan website can be found at \href{https://4chan.org}{4chan.org}.

The names 4chan as well as related names, marks, emblems and images are registered trademarks of their respective owners.

Please be aware 4chan may serve offensive and/or illegal images and texts, and it is the sole responsibility of the user to vet and remove material that may be illegal in their jurisdiction. Images are not downloaded via 4TCT, but links to them can be extracted from the data resulting from 4TCT. Texts are downloaded without any processing from the tool, and therefore may contain illegal material, including but not limited to hateful speech, copyrighted texts, calls to violence, and other potentially illegal texts.

The author and GESIS shall have unlimited liability for damages resulting from injury to life, limb or health and in the event of liability under the German Product Liability Act (Act on Liability for Defective Products). This shall also apply in the event of a breach of so-called cardinal obligations, i.e. obligations the breach of which would jeopardize the purpose of the contract and the performance of which the user as contracting party may therefore reasonably rely on. Otherwise, the author and GESIS shall only be liable for intent and gross negligence.

\section{Introduction}
4chan provides an API that allows querying of the content: textual, metadata, and images, from the servers which host 4chan (a.4cdn.org, i.4cdn.org and s.4cdn.org). The 4chan text creation tool (4TCT) utilises this API to collect the textual data and metadata from 4chan's a.4cdn.org server. 

We intentionally have not implemented image download functionality as tools already exist to do this (see Section \ref{sec:existingTools},) and to do this there likely is a burden to vet the legality for storage of these images as part of a dataset, which is not feasible at this time (especially if the data is provisioned as part of a dataset). 

\subsection{Related Work}
\label{sec:relatedworks}
\label{sec:sociology}
4chan has been of interest to a section of sociological research due to its participation in multiple political or cultural movements such as Gamergate \citep{nikkila_vivian_2021}, Pizzagate \citep{tuters_post-truth_2018}, or the Trump presidency \citep{nagle_kill_2017, hine_kek_2017, merrin_president_2019}. It also has been studied as a hub of far-right content \citep{mittos_and_2020}, including particular foci on the infamous /pol/ and /b/ boards \citep{bernstein_4chan_2011, colley_challenges_2022, jokubauskaite_generally_2020, baele_variations_2021} and also the use of memes within these movements \citep{tuters_they_2020}. It has also been used in computer science fields as a source for collection of Antisemitic and Islamophobic text and images for training models to detect such data \citep{gonzalez-pizarro_understanding_2023}.

We hope that with the release of 4TCT to the public, with its ability to monitor either particular boards or the entire site, will enable researchers to further study both this website and these topics with greater ease.
\subsection{GESIS' Digital Behavioural Data Project}
\label{subsec:DBD-SDP}
GESIS supports researchers in the social sciences throughout the research cycle by consulting the planning of studies and data collection, sourcing and acquisition of data, the processing and analysis, and archival and distribution of findings and results. Further details of this support can be found at \href{https://www.gesis.org/en/services}{https://www.gesis.org/en/services}.

GESIS plans to explore compiling and providing a dataset based on the data collected by this tool via a project that contributes to the study of social media and other digital behavioral data: \href{https://www.gesis.org/en/institute/digital-behavioral-data}{Digital Behavioural Data} (\href{https://www.gesis.org/institut/digitale-verhaltensdaten}{in German}), the datasets resulting from this project can be found \href{https://www.gesis.org/en/services/finding-and-accessing-data/digital-behavioral-data-datasets}{here}. GESIS has not yet decided whether to  collect data from 4chan, as the team is assessing the legal situation for collection, storage, cleaning and hosting data from 4chan prior to commencement.
\subsection{Existing Tools and Datasets of 4chan}
\label{sec:existingTools}
While there are multiple repositories and tools archiving \citep{bibanon_bibliotheca_2023} or for scraping 4chan \citep{bibanon_basc-archiver_2018}, this seems to be the first resource to focus on textual data collection across the entirety of 4chan. We note that multiple authors have individually collected data from 4chan for analysis, and others cite preexisting datasets.

There are many open-source resources on Github for interfacing with 4chan, including tools for scraping both images and text from 4chan \citep{issun_gchan_2023}, specifically images from either a particular board \citep{exceen_4chan-downloader_2023}, or text and images from a single \citep{devore_4chan_2023} or multiple specific threads \citep{gary_archive-chan_2023}, boards \citep{sychra_4chan-b-scraper_2022} or even individual posts \citep{denny_4chan-dl_2022}.

There are also tools that augment or present data specifically from 4chan, for example automatically downloading pastebin links \citep{woodenphone_4chan_2018}, or visualisations and UI for data collection \citep{bstrds_4chdm_2020, devix71_4chandownloader_2020}.

Most of these are run as command line or Python package (or equivalents in other languages) utilities, and as such we aim to provide a containerised tool that is more user-friendly for setting up data collection tasks such as collecting data from a board.

\section{4TCT}
\label{sec:functionality}
\subsection{Technical Information}
4TCT is currently hosted in a Github repository: \href{https://github.com/jhculb/4TCT}{https://github.com/jhculb/4TCT}, and is released under a AGPL v3.0 license. Please note that this tool may collect personal and/or illegal material that has been posted to 4chan, if you are compiling a database with this tool it is requested you comply with the highest ethical standards. 

The tool is designed for fairly simple use with a minimum level of Python or Docker knowledge required. A few examples illustrating usage have been provided. To customise the Docker runtime requires some editing of code. No tools for processing or loading the data are included.

The code has been formatted to comply with \href{https://peps.python.org/pep-0008/}{PEP8}, and further technical information about the project can be found in the included \href{https://github.com/jhculb/4TCT/blob/main/readme.md}{readme}. 
\subsection{Algorithm}
To run 4TCT, in the initialisation one may provide the requester class the boards one would like to monitor or exclude from monitoring. 

Once this has been specified the tool enters the main loop and will first check the default storage path to find data from previous runs (to allow for crash handling to prevent data duplication). The tool will then request the threads currently on each of the boards being monitored, identifying new, live (still active since last observation) and 'dead' threads (ones are no longer accessible, due to them having moved off the end of the board). Following this it will then iterate through each thread on each board and collect and store the data in .json files. 

Once it has downloaded the data from each thread it then jumps back to the start of the loop, by checking the threads that are currently on each board.
\subsection{Data Storage}
All data from 4TCT is stored in a directory created at runtime on the same level as /src/ called /data/. 

At runtime, two subdirectories are created: /data/logs/ and /data/saves/(date), for the output from logging and the raw data respectively. 

For each time 4TCT is invoked, two files will be created in /data/logs/, info\_log(timestamp).log and debug\_log(timestamp).log, which will be dynamically updated while the program executes. A blocking read from a text reader (e.g. Microsoft Word) will cause the logger to be unable to read from these files and may cause an error or crash.

The raw data collected from 4chan is stored in the subdirectory /data/saves/(date)/, two subdirectories are created, /data/saves/(date)/threads/, and /data/saves/(date)/threads\_on\_boards. 

Threads on boards contain (board name)(timestamp).json files containing the output from a API call to a.4cdn.org/(board)/threads.json, which contains the threads contained on the board at the time the program was first run. These are updated and the timestamp is renewed each loop.

Threads contains subdirectories, one for each board that has been selected, or previously collected from that day. Inside these there are .json files with file name (thread id)\_(timestamp).json. This contains the raw data from each call to a.4cdn.org/(board)/thread/thread\_id.json, which is a dictionary containing an list "posts" of dictionaries which contain all the raw data that the 4chan API provides, which depending on the post, may include:
\begin{itemize}
    \item Post ID
    \item Time of post
    \item Name of Poster
    \item Textual content (in raw format, including html tags)
    \item (optional field) Attached filename
    \item (optional field) Attached file extension
    \item (multiple optional fields) Image properties (height, width, ...)
    \item (multiple optional fields) File properties (md5, filesize, ...) 
\end{itemize}

\subsection{Future Development Avenues}
\label{subsec:futurework}
4TCT was adapted from a personal project, and as such may benefit from refactoring out design choices made before the final structure and use case was understood. These include: 

Removal of multithreading, which was implemented due to an early architecture change during a migration from a monolithic to a distributed architecture, 

Refactoring out dynamic elements such as public functions to monitor the boards, which was implemented as the early model was for a web-based application,

Removal of date-based data separations, implemented as it was not expected that (as a personal project,) collection would not be running constantly,

Changing the timing of requests for threads on a board to occur directly prior to collection of the threads themselves, to collect the most recent updates to the board at each iteration,

Packaging the code as a Python module and publishing on Pypi or a similar repository, to increase the reach and reduce the barriers of entry for the code,

And the implementation of a testing suite, type hints, and other code quality improvements to increase the robustness and maintainability of the tool.

For researchers interested in gathering both images and text, other tools mentioned in Section \ref{sec:existingTools} exist to gather image data. However extending 4TCT to gather images could be done through extracting the data for the image filename and extension, formatting this information, then requesting the image from the i.4cdn.org endpoint through the 4chan API. We have not implemented this functionality as we understand vetting the copyright, and legality, of the images is not a trivial task and the contribution to GESIS' DBD project (see Section \ref{subsec:DBD-SDP}) does not require this functionality at this time. 

Continued development of this tool is welcomed by the author, and can be done via pull requests to the Github repository.

\section{Conclusion}
\label{sec:conc}
In this technical report, we have highlighted the sociological interest of 4chan and identified a deficit in current tools' abilities to easily capture a large amount of textual data, which is currently particularly desirable as training language models require a large amount of data.

We then introduced the 4TCT tool, have briefly given a description of its algorithm and future development avenues, including its potential use in the GESIS Digital Behavioural Data program - providing a large dataset of textual data from 4chan.

We hope the release of this tool will enable individual researchers and teams to more easily collect textual data for analysis of the 4chan community. 
\section*{Acknowledgements}
This work was done during the research time with funding supplied by the Federal Ministry of Education and Research under the KB-Mining - Kompetenznetzwerk Bibliometrie project.

We thank \href{https://orcid.org/0000-0002-6656-1658}{\includegraphics[scale=0.06]{orcid.pdf}Philipp Mayr} and \href{https://orcid.org/0000-0002-4504-5144}{\includegraphics[scale=0.06]{orcid.pdf}Dimitar Dimitrov} for motivating and supporting the 4TCT project.
\bibliographystyle{unsrtnat}
\bibliography{references}  
\end{document}